\documentclass[preprint]{vgtc}               

\ifpdf
  \pdfoutput=1\relax                   
  \usepackage{graphicx}                
  \DeclareGraphicsExtensions{.pdf,.png,.jpg,.jpeg} 
\else
  \usepackage{graphicx}                
  \DeclareGraphicsExtensions{.eps}     
\fi%

\graphicspath{{figures/}{pictures/}{images/}{./}} 

\usepackage{microtype}                 
\PassOptionsToPackage{warn}{textcomp}  
\usepackage{textcomp}                  
\usepackage{mathptmx}                  
\usepackage{times}                     
\usepackage{cite}                      
\usepackage{tabu}                      
\usepackage{booktabs}                  

\newcommand{\pr}{Plotly-Resampler}
\usepackage{listings}
\usepackage[dvipsnames]{xcolor}
\usepackage[frozencache=true,cachedir=minted-cache]{minted}
\usepackage{enumitem}
\usepackage{url}
\usepackage[title]{appendix}
\usepackage{subfig}
\usepackage{rotating}
\usepackage{tikz}
\usepackage{newfloat}
\usepackage{caption}
\usepackage{balance}
\usepackage[breaklinks=true]{hyperref}

\definecolor{codegreen}{rgb}{0,0.6,0}
\definecolor{codeblue}{rgb}{0,0,0.6}
\definecolor{codegray}{rgb}{0.5,0.5,0.5}
\definecolor{codepurple}{rgb}{0.58,0,0.82}

\usepackage{ascii}

\lstdefinestyle{mystyle}{
    language=Python,
    commentstyle=\color{codegray},
    keywordstyle=\color{codegreen},
    numberstyle=\tiny\color{codeblue},
    stringstyle=\color{codepurple}\scriptsize\asciifamily,
    basicstyle=\scriptsize\asciifamily,
    breaklines=true,                 
    captionpos=b,                    
    keepspaces=true,                 
    numbers=left,                    
    showstringspaces=false,
    tabsize=2,
    morekeywords={True, False},
    basewidth={.55em}
}
\onlineid{1021}

\vgtccategory{System or tool}

\vgtcinsertpkg

\preprinttext{To appear in an IEEE VGTC sponsored conference.}

\title{Plotly-Resampler: Effective Visual Analytics for Large Time Series}

\newcommand*\samethanks[1][\value{footnote}]{\footnotemark[#1]}

\author{
    Jonas Van Der Donckt 
    \thanks{contributed equally} 
    \thanks{e-mail: jonvdrdo(dot)(lastname)(at)ugent(dot)be} 
    \and Jeroen Van Der Donckt  
    \samethanks[1]
    \and Emiel Deprost  
    \and Sofie Van Hoecke  
}
\affiliation{\scriptsize IDLab, Ghent University - imec, Belgium}

\teaser{
  \centering
    \includegraphics[width=\linewidth]{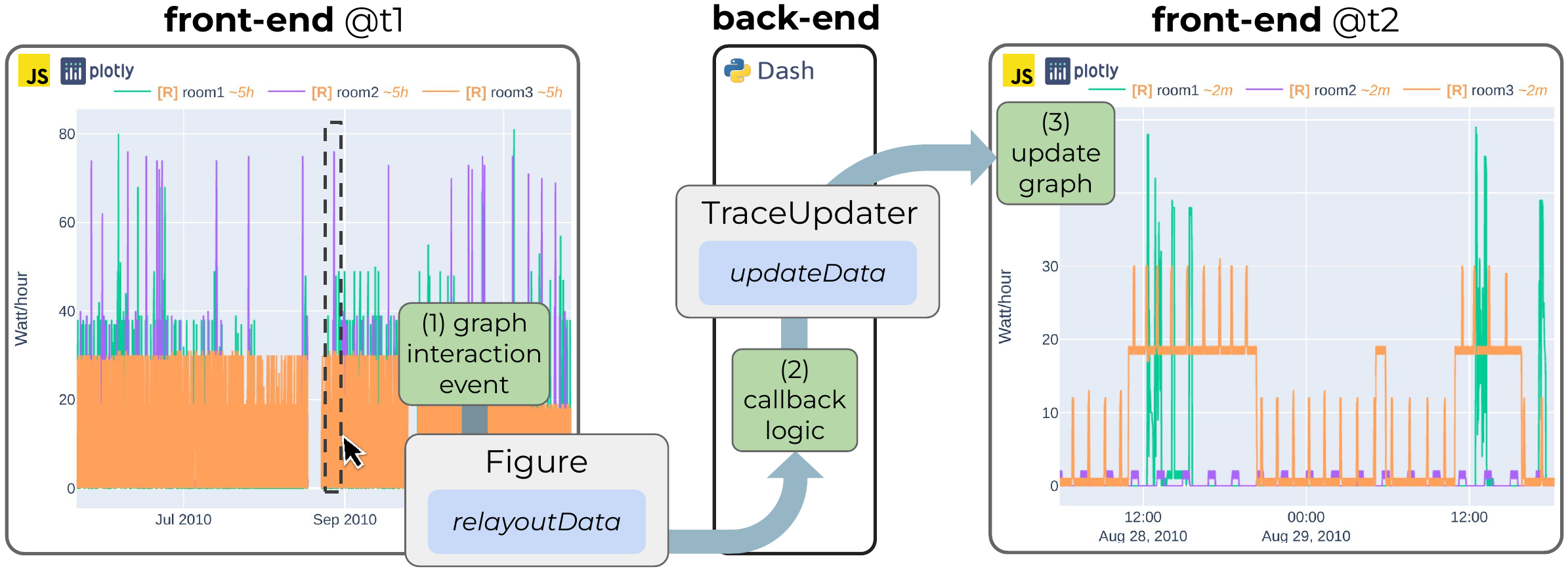}
    \vspace*{-5mm}
  \caption{
High-level overview of \pr{}'s dynamic aggregation functionality (see \href{https://raw.githubusercontent.com/predict-idlab/plotly-resampler/fbd932bbb96800136fbc72ab95d9676373395be3/docs/sphinx/_static/basic_example.gif}{gif}). A front-end graph interaction event (i.e, zoom event) (1) triggers a callback, sending the layout-change (i.e., \textit{relayoutData}) to the back-end. In the Dash back-end (2), the \textit{relayoutData} is processed to perform data aggregation for the new regions of interest. Finally, only the to-be-updated data (i.e., \textit{updateData}) is sent to the front-end, on its end triggering a graph update (3). The presence of an \textcolor{Orange}{\textbf{[R]}} legend prefix indicates that an aggregation of the data is shown. The \textcolor{Orange}{$\sim$\textit{$<$time$>$}} suffix in the legend represents an estimate of the aggregation bin size.
  }
  \label{fig:teaser}
}

\abstract{
Visual analytics is arguably the most important step in getting acquainted with your data. This is especially the case for time series, as this data type is hard to describe and cannot be fully understood when using for example summary statistics.
To realize effective time series visualization, four requirements have to be met; a tool should be (1) interactive, (2) scalable to millions of data points, (3) integrable in conventional data science environments, and (4) highly configurable.
We observe that open source Python visualization toolkits empower data scientists in most visual analytics tasks, but lack the combination of scalability and interactivity to realize effective time series visualization. 
As a means to facilitate these requirements, we created \href{https://github.com/predict-idlab/plotly-resampler}{\pr{}}, an open source Python library.
\pr{} is an add-on for Plotly's Python bindings, enhancing line chart scalability on top of an interactive toolkit by aggregating the underlying data depending on the current graph view.
\pr{} is built to be snappy, as the reactivity of a tool qualitatively affects how analysts visually explore and analyze data.
A benchmark task highlights how our toolkit scales better than alternatives in terms of number of samples and time series.
Additionally, \pr{}'s flexible data aggregation functionality paves the path towards researching novel aggregation techniques.
\pr{}'s integrability, together with its configurability, convenience, and high scalability, allows to effectively analyze high-frequency data in your day-to-day Python environment.
} 

\CCScatlist{
  \CCScatTwelve{Time series}{Visual analytics}{Python}{Dash}
  \CCScatTwelve{Plotly}{Open source}
  {}{}
}

\nocopyrightspace

\begin{document}



\maketitle

\section{Introduction}  

%

Data wrangling is the process of iterative data exploration and transformation, enabling downstream tasks such as modeling and data analysis. In the data wrangling process, and even in the whole data science pipeline, visual analytics has proven to be a crucial component. For example, visualizations can be utilized to assess the quality of your processing, validate the alignment of your data, obtain insights, and even analyze model predictions. After all, the human eye has frequently been advocated as the ultimate data mining tool~\cite{lin_visualizing_2005}.


However, effective visualization of time series remains challenging as this data type is less intuitive to grasp compared to for example images or text. Both the temporal and multivariate aspect of time series data should be captured in meaningful visualizations. It has been shown that for most time series analysis tasks, simple visualizations (e.g., line-based charts) are sufficient, compared to more complex or use-case specific approaches~\cite{aigner2007visualizing}.

Due to decreasing costs in both storage and sensors, large time series data gets more common. We observe that the current Python visualization landscape struggles to effectively handle such large datasets~\cite{veljanoski_interactive_2020}. More specifically, some tools lack graph interactivity, which is necessary to effectively explore time series, whereas other interactive tools become slow or unresponsive with such large quantities of data.
To tackle this problem we created \pr{}, an add-on for Plotly's Python bindings, enabling interactive visualizations of large sequences. \pr{} complies with the visual information seeking mantra: \textit{``Overview first, zoom and filter, then details-on-demand''}~\cite{shneiderman_eyes_2003}. Correspondingly, \pr{} is built to be snappy, as the speed of data retrieval qualitatively affects how analysts visually explore and analyze their data~\cite{sye-min_chan_maintaining_2008}.

The contribution of this paper is threefold:
\begin{itemize}[itemsep=-1.5pt]
    \item We \textbf{introduce \pr{}}, an open source Python toolkit that enables interactive and scalable time series visualization. This is achieved by performing under-the-hood data aggregation, depending on the current graph view. 
    \item We \textbf{demonstrate} the \textbf{practical usefulness} of \pr{} for data wrangling and downstream tasks with a real-world use case and show the minimal code overhead of our package.
    \item We \textbf{position and benchmark} \pr{} against alternatives and highlight its strengths and limitations.
\end{itemize}

\section{Related work}\label{sec:related}
According to the Kaggle 2021 survey, (IPython-)notebook-based environments are the go-to tools for data-scientists~\cite{noauthor_kaggles_2021}.
Such a notebook-based format drives exploration, which is crucial in every step of the data science process. Their form of interactive computing enables users to execute code, observe, modify, and repeat in an iterative conversation between analyst and data~\cite{perkel2018jupyter}. 

Based on the observations of Bikakis~\cite{Bikakis2018} and Perkel~\cite{perkel2018jupyter}, we list four requirements for an effective time series visualization tool;
\begin{enumerate}[itemsep=-1.5pt]
    \item \textit{Provide interactivity}: To enable effective exploration, and thus comply with the information seeking mantra, visualizations need to support functionalities like zooming, panning, showing information on hover, and (de)selecting specific traces~\cite{shneiderman_eyes_2003, aigner2007visualizing}.
    \item \textit{Scalable to large datasets}: Visualizations should not become slow or unresponsive with large data, where millions of observations from multiple modalities are common.
    \item \textit{Integrable in conventional data-science environments}: The toolkit should be easy to integrate with existing tools and workflows, for example IPython based notebooks running on a remote server.
    \item \textit{Configurable and convenient}: Having a tool that is both easy to use and highly configurable enables broad applicability. For example, an effective tool should conveniently allow visualizing multivariate data on a shared time axis. 
\end{enumerate}

In the last two decades, a lot of research has been done on time series visualization techniques~\cite{kincaid2010signallens, walker2015timenotes, lin_visualizing_2005, zhao2011exploratory, cho2014stroscope, morrow2019periphery, stopar2018streamstory, lin2004viztree}. This work focused on describing and comparing time series visualization approaches. 
However, when we position this research with regard to practical applicability, a common trend is observed;
these contributions do not seem to focus on code availability nor integration with either visualization packages or data wrangling environments. This severely limits their applicability, and also makes it hard to reproduce results or to assess other aspects such as scalability and interactivity. 
Only Stopar et al.~\cite{stopar2018streamstory} and Kincaid et al.~\cite{kincaid2010signallens} provided benchmarking information about their methodology, whereas only Morrow et al.~\cite{morrow2019periphery} made their code publicly available.

\begin{table}[b]
\centering
\vspace*{-3mm}
\caption{Requirement assessment for effective visualization of large time series, applied to popular open source libraries.}
\begin{tabular}{@{}llcccc@{}}
\toprule
  &                 & Bokeh & Plotly    & Matplotlib    & Holoviews \\ \cmidrule(l){2-6} 
1 & Interactivity   & +     & +         & -             & + \\
2 & Scalability     & -     & -         & $\pm$          & $\pm$ \\
3 & Integrability   & +     & +         & +             & +      \\
4 & Configurability & +     & +         & +             & + \\ \bottomrule
\end{tabular}
\label{tab:req_mapping}
\end{table}

Given the lack of accessible research outcomes on the one hand and the Kaggle survey results on the other hand, we limit our comparison to open source Python toolkits. Table~\ref{tab:req_mapping} shows that none of those selected toolkits meets all the aforementioned requirements. We observe that there is a consistent trade-off between interactivity and scalability. Matplotlib~\cite{Hunter2007matplotlib} is somewhat scalable to larger data sizes, but has limited interactivity. 
Both Bokeh~\cite{bokehpython} and Plotly~\cite{plotlypython} provide extensive interactivity by using a JavaScript front-end. However, loading large quantities of data in this front-end hinders responsiveness, thus restricting the scalability of these two libraries.
Finally, HoloViews approaches visualization by providing a set of data structures that pair your data with a small amount of metadata~\cite{stevens2015holoviews}. Those data structures are then rendered by a separate plotting system, e.g., Bokeh or Plotly, to visualize data interactively. Just as \pr{} does, HoloViews provides data aggregation functionality through its Datashader bindings. However, these are not optimized for multivariate line chart aggregation, limiting the convenience and scalability of HoloViews.

As a result, users tend to settle for suboptimal approaches such as downsampling or selecting smaller parts of the data before creating interactive visualizations. On the one hand, static data aggregation (i.e., downsampling) does not show the full data and can induce artifacts such as aliasing~\cite{mishali2011sub}. On the other hand, visualizing smaller parts of the data results in a shattered interaction process, as it is substantially harder to associate the overall data with these sporadic local inspections.

It is our first-hand experience with the above shortcomings that sparked the creation of \pr{}.


\section{Plotly-Resampler toolkit}





Visualizations are limited by the size of the canvas (i.e., pixel density of the screen).
As such, a common issue when visualizing large quantities of data is over-plotting, where multiple data points are rendered on top of each other~\cite{datashaderpitfalls, walker2015timenotes}. Furthermore, front-end rendering and responsiveness slow down significantly when more data points are portrayed.
%
These problems with visualizing large quantities of data can be tackled, to some extent, by either employing series-wise data aggregation (as \pr{} does) or using density-wise aggregation (as Datashader~\cite{datashader} does).


Datashader was developed to represent large amounts of data for which a shared density color coding is used, making it the most effective when there is a single, large, data modality (e.g., a map, a point cloud). However, when there are multiple distinct modalities in the data, each requires a distinguishable color density coding, resulting in (again) an over-plotting issue. Unfortunately, time series often consists of distinct modalities, rendering this technique unsuitable.

Alternatively, series-wise data aggregation controls over-plotting by reducing the number of rendered data points. This makes multivariate visualizations scalable at the cost of possibly inducing information loss~\cite{agrawal_challenges_2015}. However, a key insight into visualizing such large amounts of data is that not all data points are equally important. The Largest-Triangle-Three-Buckets (LTTB) aggregation technique exploits this insight to effectively select data points~\cite{steinarsson2013downsampling}.

From these observations, it is clear that series-wise data aggregation is most suitable for effective time series visualization. Next to this, in contrast with a custom plotting language (e.g., HoloViews), we opted to build upon Plotly. 
By complying with the \textit{"Don't reimplement the wheel"}-mantra, our toolkit leverages all of Plotly's functionality, i.e., convenience, integrability, interactivity \& high configurability, while at the same time resulting in a smaller, more manageable codebase. 
The realization of \pr{}'s goal, i.e., scalable data aggregation, is shown in Figure~\ref{fig:teaser}.




\begin{figure*}[tb]
 \centering 
 \includegraphics[width=\linewidth]{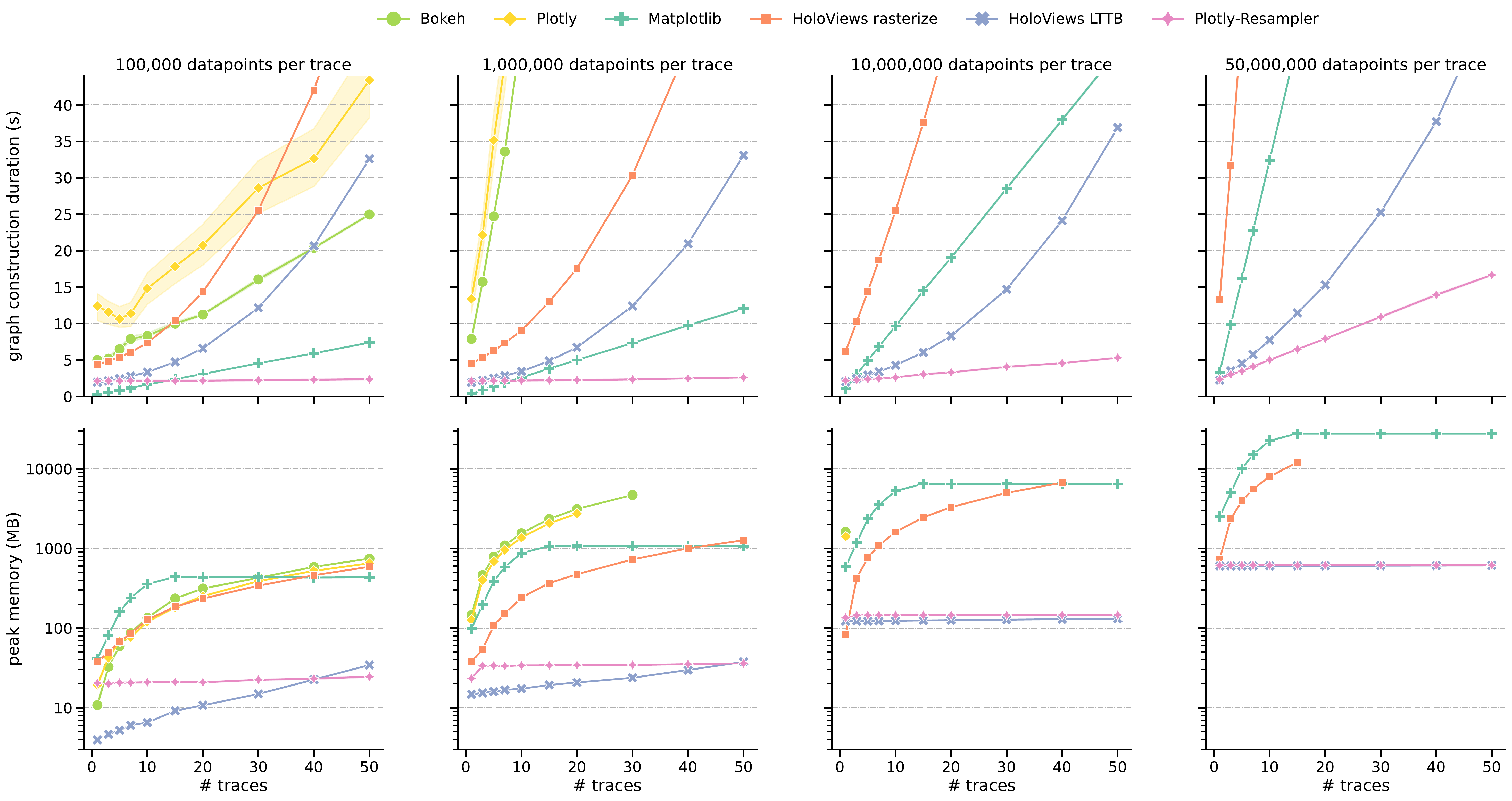}
 \vspace*{-5mm}
 \caption{Benchmark results for a line-graph visualization task (\href{https://github.com/predict-idlab/plotly-resampler-benchmarks}{code}~\cite{vanderdoncktPRBenchmarks}). The top charts display the average duration of constructing and rendering the graph. The bottom charts indicate the peak memory usage. The columns indicate the data size per signal, thus showing a trend when scaling to larger datasets.
 In each chart, the number of visualized traces (i.e., lines) is represented on the x-axis. Each toolkit configuration was benchmarked until the duration exceeded 2 minutes.
 Dynamic aggregation is realized in \texttt{HoloViews LTTB} by using \pr{}'s LTTB aggregation functionality, whereas \texttt{HoloViews} \texttt{rasterize} uses the built-in Datashader-rasterize function.
 }
 \vspace*{-3mm}
 \label{fig:benchmark}
\end{figure*}

\subsection{Features}\label{subsec:features}
With \pr{} we aim to contribute a qualitative Python package to the data visualization community for effective time series analysis.
For that reason, \pr{} is listed on \href{https://pypi.org/project/plotly-resampler/}{PyPi}, and can thus be installed via \texttt{pip install plotly-resampler}. \pr{}'s source code is available on \href{https://github.com/predict-idlab/plotly-resampler}{GitHub}~\cite{vanderdoncktPRBenchmarks}, encouraging input from the community (e.g., contributions via pull requests, issues for discovered bugs, or feature requests). For detailed information, users can consult the \href{https://predict-idlab.github.io/plotly-resampler/}{documentation} along with several code \href{https://github.com/predict-idlab/plotly-resampler/tree/main/examples}{examples} demonstrating how to apply \pr{} for various visualization tasks.
To assure \pr{}'s quality, the functionality is validated with a CI-CD testing pipeline, which
tests the toolkit's functionality as a web-app with Selenium~\cite{holmes2006automating} and validates the content of the transferred packets (triggered by graph callbacks)~\cite{seleniumwire}.

The scalability of \pr{} is realized by optimizing several aspects.
To perform data-updates, the default Dash graph component requires re-sending all graph data to the front-end, resulting in unnecessary data and computation overhead. This behavior causes an increased callback latency, thus impacting responsiveness. To overcome this problem, we created TraceUpdater~\cite{vanderdoncktTraceUpdater}, a Dash component that minimizes callback latency by only sending to-be-updated trace data to the front-end. 
Additionally, the back-end data access latency is negligible as the back-end data is stored in memory. 
Finally, code profiling allowed us to look for bottlenecks and validate that view-based operations (i.e. pass by reference) were applied to the data where possible as such operations do not induce large memory and runtime overheads.

Other features of the \pr{} toolkit are a direct consequence of its flexibility. For example, \pr{} is not limited to numeric data types, but also supports categorical and boolean series. Furthermore, there is a high configurability of the functionality, e.g., the data aggregation technique, time series gap detection, 
or how the web-app should be served. As \pr{} is built in the Plotly-Dash ecosystem, it also integrates seamlessly with Dash web applications. 



\subsection{Benchmarks}\label{subsec:benchmarks}
Given the prevalence of large and high-frequency time series, the scalability capabilities
of eligible toolkits were assessed with a line-graph visualization benchmark. 
The profiling is realized using the \texttt{VizTracer}~\cite{viztracer} package with the \href{https://github.com/gaogaotiantian/vizplugins}{\texttt{VizPlugins}} add-on. The benchmarking code is available on \href{https://github.com/predict-idlab/plotly-resampler-benchmarks}{GitHub}~\cite{vanderdoncktPRBenchmarks}, encouraging reproducibility.
We further refer to \href{https://github.com/predict-idlab/plotly-resampler-benchmarks/tree/6fda4b25f42bb034168f345a97a785562350e2cb/reports}{this README}~\cite{prbenchmarksprocedure} for detailed information regarding the benchmarking procedure.


\begin{figure}[tb]
 \centering 
 \includegraphics[width=\columnwidth]{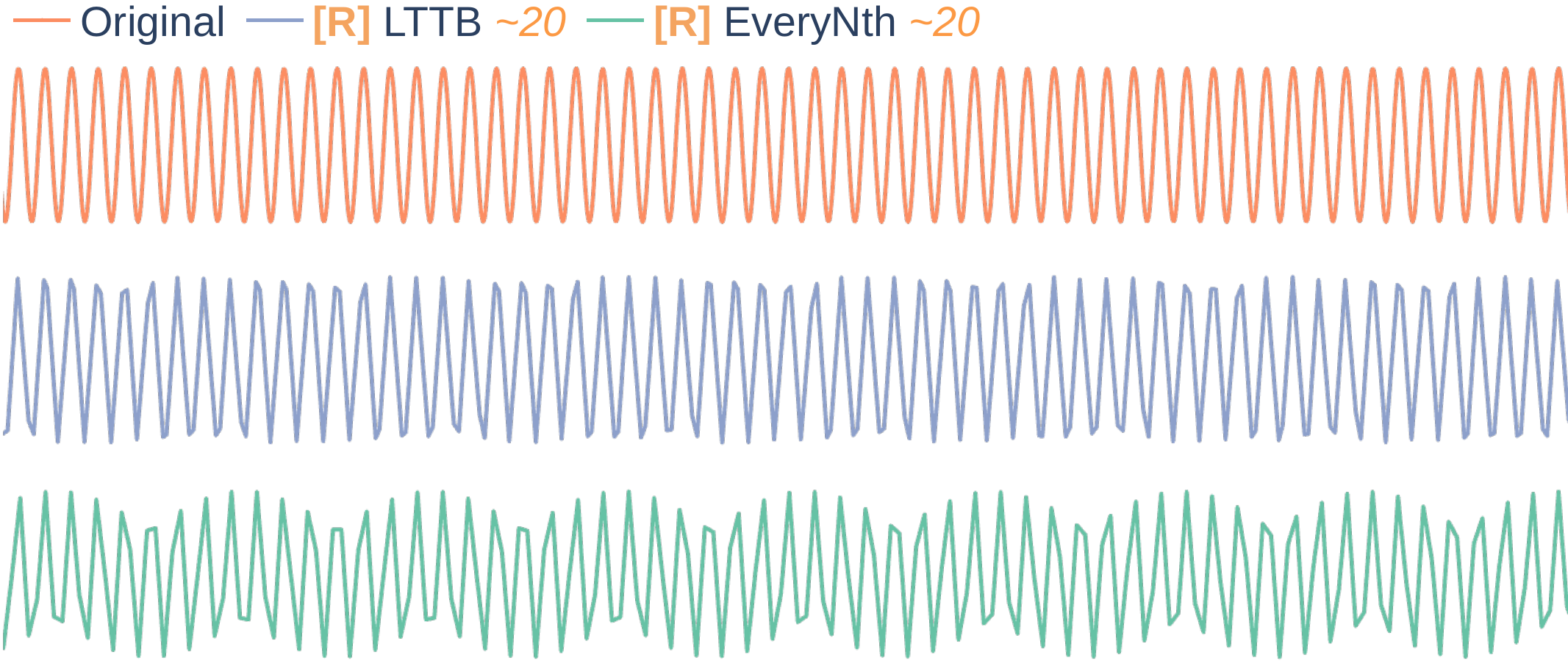}
 \caption{Influence of data aggregation method on aliasing.
 The middle graph uses the \texttt{LTTB} aggregation algorithm, whereas the bottom graph employs the naive \texttt{EveryNth} algorithm.
 }
  \vspace*{-4mm}
 \label{fig:aliasing}
\end{figure}

Figure~\ref{fig:benchmark} depicts the profiling results. 
The top charts represent the total time to construct and render the visualization.
As expected, the graph construction time increases with the number of data points (i.e., columns) and number of traces (i.e., x-axis), however not all at the same pace.
\texttt{\pr{}} clearly scales better in terms of data points and number of traces. \texttt{Bokeh} and \texttt{Plotly} (without any aggregation functionality) scale the worst, as they are heavily affected by the dataset size.
The visualization time of HoloViews-based approaches scales exponentially in terms of the number of visualized traces (i.e., x-axis), rendering them unsuitable for large multivariate visualizations. 


The bottom charts indicate the peak memory usage. We observe that both \texttt{\pr{}} and \texttt{HoloViews LTTB} scale better. These two approaches use less than 700 MB for the largest configuration (right column), whereas \texttt{Matplotlib} and \texttt{HoloViews rasterize} exceed 10 GB.

When dealing with more than 10,000,000 samples per series and more than 10 traces, \texttt{\pr{}} emerges as the only viable toolkit regarding graph construction time and memory usage.

\begin{figure*}[!htb]
    \centering
    \includegraphics[width=\linewidth]{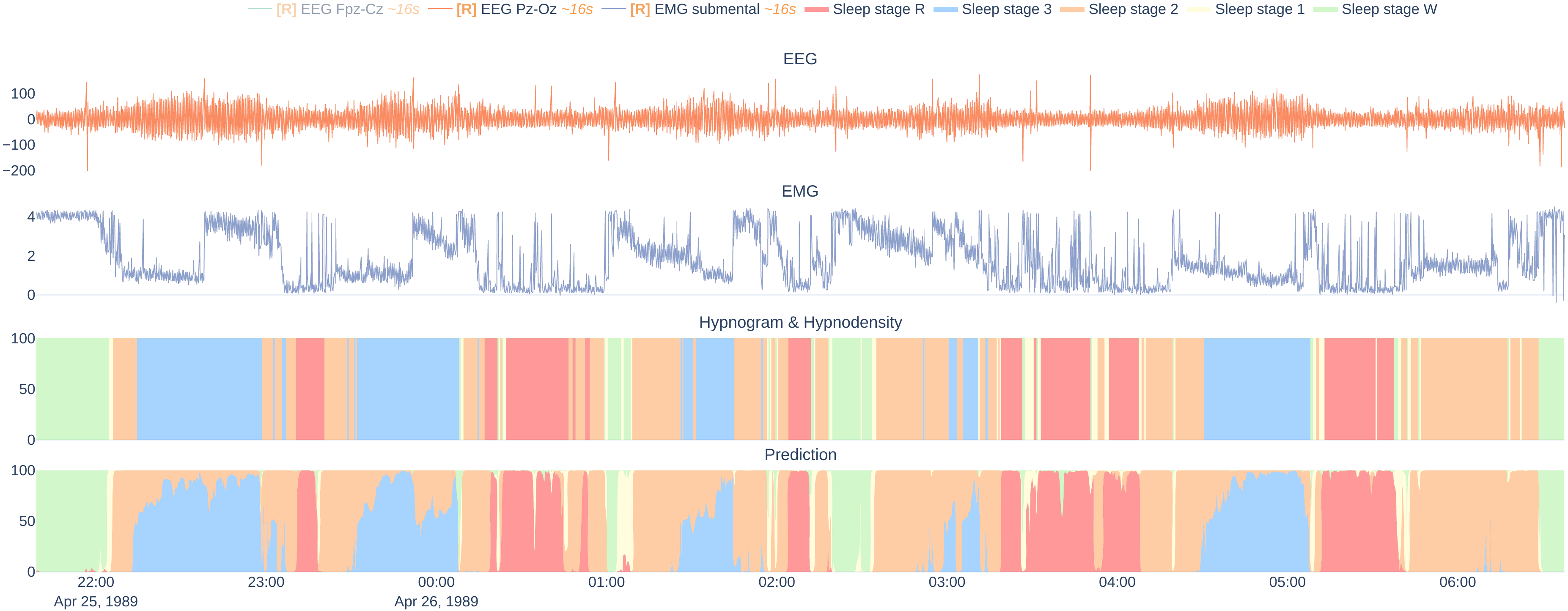}
    \vspace*{-6mm}
    \caption{Sleep stage classification example, demonstrating the joined visualization of raw sensor data with probabilistic predictions. The first two rows show the raw data, i.e., EEG signals which are sampled at 100Hz, and an EMG signal which is sampled at 1Hz, respectively. The third row shows the sleep stages, also called a hypnogram on the right y-axis (black line), and the hypnodensity, i.e., a probabilistic hypnogram, on the left y-axis. The final row shows probabilistic hypnodensity predictions of a machine learning pipeline, empowered by tsflex~\cite{van2022tsflex}.}
    \label{fig:sleep_dashboard}
    \vspace*{-4mm}
\end{figure*}

\subsection{Limitations}

Data aggregation is a form of resampling, rendering it susceptible to downsampling artifacts, in particular aliasing~\cite{mishali2011sub}.
Aliasing is more prevalent when a naive aggregation algorithm, such as \texttt{EveryNth} is used (where data points are aggregated by sampling with a fixed interval $N$). 
More advanced aggregation methods seem to be less prone to such artifacts.
For example, LTTB samples data points in bins with the goal of maximizing the triangular surface of surrounding bins~\cite{steinarsson2013downsampling}. 
As an illustration, Figure~\ref{fig:aliasing} highlights the influence of the aggregation method on aliasing artifacts. We observe that for the shown region of interest, both methods display artifacts. As expected, \texttt{EveryNth} aggregation (bottom graph) results in a more distorted view than \texttt{LTTB} (middle graph).
\pr{} attempts to minimize the aliasing issue by employing an efficient heuristic of LTTB as default aggregation algorithm and by indicating when data aggregation is performed. This indication is realized by displaying the \textcolor{Orange}{\textbf{[R]}} legend prefix as well as an estimate of the aggregation bin-size as legend suffix.

\pr{} supports various data types, such as numeric, categorical, and boolean time series data. However, it cannot consider all aspects involved when visualizing time-oriented data, e.g., time-bound text data. Aigner et al.~\cite{aigner2007visualizing} indicated that such different types of time-oriented data can only be visualized with dedicated methods. Hence, we argue that supporting such visualizations is out of scope for this toolkit.

\pr{} utilizes the \href{https://pandas.pydata.org/pandas-docs/stable/reference/api/pandas.Series.html}{\texttt{pandas.Series}~\cite{reback2020pandas}} data container for storing high-frequency time series. This induces the limitation that all visualized data must fit in memory. 


\section{Toolkit usage and example use cases}


Using \pr{} requires minimal code overhead. 
To add aggregation functionality to a Plotly figure, end-users only need to wrap their Plotly figure with the \texttt{FigureResampler} decorator and call \texttt{show\_dash()} on that object. 
A minimal code example as a \href{https://gist.github.com/jonasvdd/828ea56dbfef21e9999392234c38a8af}{GitHub gist} illustrates this usage~\cite{pr_usage}. 


\subsection{Use case: sleep staging model analysis}\label{subsec:dashboard_example}


Figure~\ref{fig:sleep_dashboard} highlights how \pr{} can be used for sleep (polysomnography) data analysis~\cite{berry2012rules}.
\pr{} allows us to visualize a full night recording of electroencephalography (EEG), and  electromyography (EMG) data with a machine learning model its predictions in one graph. 
From Figure~\ref{fig:benchmark} we observe that no other tool is capable of serving the scalability to interactively visualize such large quantities of multivariate data.
\pr{} gives a global overview in which we can see that the model probabilities are well-adjusted. Furthermore, we can obtain insights by zooming in on regions of interest, e.g., mispredictions, sudden changes in the raw data.

\section{Conclusion}
%
In this paper, we present \pr{}, an open source Python package enabling effective time series visualization. 
An effective visualization toolkit should be (1) interactive, (2) scalable, (3) integrable, and (4) configurable. We observe that no existing Python package meets these four requirements, as there is a consistent trade-off between interactivity and scalability. \pr{} tackles this trade-off by adding scalability to a toolkit that already meets requirements 1, 3, and 4 (namely Plotly).
To achieve this scalability, \pr{} 
separates the visualization in a back-end and front-end. The back-end stores all the data, whereas the front-end shows an aggregated view of this data. The interactivity, i.e., dynamic data aggregation, is realized through optimized callbacks between the front-end and back-end.

Benchmarks indicate that \pr{} significantly outperforms existing alternatives when scaling to large, multivariate datasets.
Moreover, we also highlight the applicability and convenience of \pr{} with a real-world use case.

\pr{}'s flexible architecture enables it to serve as a platform for researching time series visualization approaches. In particular, it can be used to compare and research data aggregation techniques. 


With \pr{} we aim to contribute a qualitative Python package to the data visualization community. Our goal is to enable effective time series visualization in your day-to-day Python environment.


\acknowledgments{Jonas Van Der Donckt (1S56322N) and Emiel Deprost (1SA4321N) are funded by a doctoral fellowship of the Research Foundation – Flanders (FWO). 
Part of this work is done in the scope of the imec.AAA Context-aware health monitoring project.
We thank Michael Rademaker for reviewing the manuscript.
}

\bibliographystyle{abbrv}

\bibliography{template}

\newpage
\onecolumn

\end{document}